# Can Large Language Models Estimate the Likely Technological Impact of Scientific Research? Evidence from Biotechnology

Zhiyu Tao [a], Mike Thelwall [b], Danqun Zhao [a], Yi Bu [a*]

[a] Department of Information Management, Peking University, Beijing, China

[b] School of Information, Journalism and Communication, The University of Sheffield, Sheffield, UK

**Abstract:** This study provides large-scale empirical evidence about whether large language model (LLM)-based assessments of the likely technological impact of scientific publications are associated with their patent-based impact. Using Lens.org, we constructed a biotechnology dataset of 107,615 papers published between 1980 and 2015 with their patent citation counts. We used GPT-5-mini, Qwen-Flash, and DeepSeek-V3.2 to estimate papers' technological impact and technology relevance, benchmarking model outputs against log-transformed patent-citation counts through correlation analysis, distributional alignment, and lift-over-random measures. The results show that LLMs extract meaningful signals associated with technological impact from scientific text, although there are substantial differences at the meso-level such as between research fields, journals, and countries. LLMs also outperform random selection in screening for papers with high observed patent impact. Overall, this study identifies the strengths, contexts, and limitations of LLM-based technological impact evaluation and highlights its potential to support large-scale screening for scientific work with downstream technological value.

## 1. Introduction

Large language models (LLMs) are reshaping the practice of scientific research and are becoming deeply embedded in innovation processes (Shao et al., 2025; Zhang, Khan, et al., 2025). Their applications now span multiple stages of research, including hypothesis generation (Sourati & Evans, 2023),experimental design (Boiko et al., 2023), materials discovery (M. Bran et al., 2024), academic writing (Liang, Zhang, Wu, et al., 2025), and scientific funding application (Qian et al., 2026). Against this background, understanding the appropriate and effective use of LLMs has emerged as a critical question for scholars, institutions, and policymakers.

Despite the rapid diffusion of LLMs, evidence about their effective use and limitations remains limited. Studies have examined research quality assessment (Thelwall, 2025g), peer-review assistance (Zhuang et al., 2025), and disruptive technology identification (Huang et al., 2025), but have largely treated publications and patents as separate analytical objects. Whether LLMs can assess science–technology linkages therefore remain underexplored. Technological impact is difficult to observe promptly: patent citations typically emerge later than paper-to-paper citations, and research may inform technology without receiving patent citations. If LLMs can infer technological potential from papers' textual and bibliographic features, they could support technology foresight and science policy, but this ability requires empirical validation.

* Corresponding author: Yi Bu, E-mail: buyi@pku.edu.cn

We examine whether LLMs can assess the technological impact of scientific papers from textual and bibliographic features. Two questions guide the analysis: (1) To what extent do LLM-based scores align with observed patent-based impact? (2) Under what conditions are these assessments more or less informative?

## 2. Literature review

### 2.1. LLM-based practices in academic evaluation

As scientific publishing grows, identifying and assessing research outputs becomes increasingly difficult. LLMs can process large volumes of text and capture semantic relationships, prompting applications in academic writing support, research quality assessment, peer-review support, and scholarly evaluation (Liang, Zhang, Codreanu, et al., 2025; Thelwall, 2025d; Thelwall & Ehsan, 2026; Thelwall & Yang, 2025).

Existing studies have mainly examined LLM-based scientific evaluation from two perspectives. The first is content-based research quality assessment. This stream investigates whether LLMs can evaluate the originality, rigor, contribution, and potential impact of academic outputs (Thelwall, 2025d, 2025f; Thelwall & Ehsan, 2026; Thelwall & Yang, 2025). These findings suggest that LLMs can capture research quality from textual information. However, their performance is far from uniform. Evidence from evaluations of books, journal rankings, departments, and cross-disciplinary articles indicates that the usefulness of LLM-based assessment varies substantially across tasks and objects (Langfeldt et al., n.d.; Thelwall & Cox, 2025).

The second perspective concerns peer-review support. Large-scale analyses show overlap between LLM-generated and human reviews, with models sometimes providing additional suggestions (Idahl & Ahmadi, n.d.; Liang et al., 2024; Sun, 2025). LLMs have also been used to predict review outcomes (Thelwall & Yaghi, 2025). However, they may overemphasize implications, data volume, and writing while underemphasizing novelty and ablation experiments (Liang et al., 2024).

Some other studies seek to examine factors influencing the performance of LLMs, and offer methodological evidence to improve it. Empirical evidence suggests that model choice (e.g. ChatGPT, Gemini, DeepSeek) (Thelwall, 2025f), input text (e.g., title, abstract, or full text) (Thelwall, 2025e), prompt design (Thelwall, 2026a), repeated scoring (Thelwall, 2025d; Thelwall & Ehsan, 2026), and output format (Thelwall & Yang, 2025) can influence evaluation results, but the performance varies across different contexts. For instance, GPT-4o, GPT-4o-mini, Gemini, and Gemma generally show positive correlations with expert-based evaluations (Thelwall, 2025f), but their relative performance varies across tasks, input materials, and disciplines. Input based on titles and abstracts may outperform title-only or full-text inputs in some settings (Thelwall, 2025e). Prompting strategies, including chain-of-thought prompting, can improve performance in particular tasks, while repeated scoring and score averaging tend to reduce stochastic variation and improve stability (Thelwall et al., 2026). These findings provide important guidance for designing LLM-based evaluation systems.

Overall, existing research demonstrates that LLMs have meaningful potential in scientific evaluation. However, much less is known about whether LLMs can evaluate the downstream technological relevance of scientific research, which is a critical gap because technological impact depends not only on scholarly quality, but also on whether scientific knowledge can be translated, reused, and incorporated into technological invention.

## 2.2. The science-technology linkage: how scientific research informs technological innovation

### 2.2.1. Science as a source of technological innovation

Basic research can support technological innovation, economic growth, and national competitiveness (European Commission. Directorate General for Research and Innovation., 2025). Governments have consequently sought to translate scientific outputs into applications through initiatives such as China's innovation-driven development strategy (CPC Central Committee, 2016), the UK Science and Technology Framework (Department for Science, Innovation & Technology, 2025), and Horizon Europe (European Commission. Directorate General for Research and Innovation., 2025).

Translation is not automatic: discoveries must be identified, recombined, and absorbed by inventors or firms to generate technological value (Gittelman & Kogut, 2003). Identifying papers likely to inform invention is therefore relevant to technological foresight, research evaluation, and knowledge transfer.

### 2.2.2. Conceptual models of the science-technology relationship

The initial theoretical contribution to the relationship between science and technology is often credited to Vannevar Bush, who argued that scientific discovery provides the knowledge base from which later technological applications may emerge (Vannevar, 1945). Subsequent studies have developed more nuanced models of the science–technology relationship, including The Frascati model (OECD, 2002), Pasteur's Quadrant (Stokes, Donald E, 2011), and Dynamic and interactive models (Stokes, Donald E, 2011). More recent perspectives describe the science–technology relationship as part of a complex innovation system where universities, firms, governments, markets, and intermediaries jointly shape the movement of knowledge from discovery to application (Nelson, 1993).

These models explain why science matters for technology and how the domains interact, but offer limited tools for identifying technological potential in individual papers. This gap motivates approaches that detect paper-level signals of technological relevance.

### 2.2.3. Quantitative measurement of science-technology linkages

The availability of large-scale publication and patent databases has created opportunities to examine the relationship between science and technology quantitatively. This research has mainly developed quantitative measurements along three related lines.

The first is paper–patent citation analysis, the most widely adopted. Patent citations to scientific papers provide a direct way to observe how scientific knowledge is used in technological invention (Ahmadpoor & Jones, 2017). This approach has been used not only to map knowledge flows from science to technology, but also to reveal the knowledge base of specific patents, firms, or technological fields (L. Wang et al., 2020).

The second approach focuses on human or organizational linkages between science and technology. In this stream, scholars examine authors, inventors, and research teams that participate in both paper publication and patent application. These actors are often described as "gatekeepers" because they connect scientific knowledge production with technological invention (Breschi & Catalini, 2010). By constructing co-authorship or co-invention networks, researchers can identify individuals and organizations that bridge the two domains (Lissoni, 2010).

The third approach examines science–technology linkages at the topic, semantic, or knowledge-network level. This method identifies connections by comparing the content, topics, keywords, or knowledge structures of scientific publications and patent documents. Studies have used topic similarity (Motohashi et al., 2024), topic-word linkages (Xu et al., 2020), topic composition (Chen et al., 2023), semantic similarity, and knowledge-network coupling (Ba & Liang, 2021) to capture broader and more latent relationships between science and technology. For example, topic-level mapping can show how scientific topics and technological topics overlap, diverge, or coevolve over time (Xu et al., 2020), providing quantitative evidence of science–technology coevolution.

However, those quantitative approaches still have important limitations: citation-based approaches are retrospective and delayed, especially with long patent citation windows. In response to these limitations, some studies have explored machine-learning-based patent citation prediction. Such work leverages deep neural networks, multitask learning, and graph neural networks together with patent metadata (e.g., inventors, ireference counts, number of claims) to forecast the potential impact of patents (Nandi et al., 2024; Noh & Lee, 2022; Seol et al., n.d.). Meanwhile, other research treats patent citations as a complex system and models the citation process using a Hawkes process (X. Liu et al., 2017). However, existing patent impact prediction studies focus on individual patents and evaluate only the inherent technological impact of patents. Although Gao et al. (2024) extend this line of inquiry by using paper-level attributes, including scholarly citation counts and citation positions, to predict patent citations to scientific papers, their analysis focuses on artificial intelligence. The applicability of this approach to other fields remains uncertain, and extending it across domains may require additional model training and validation. This highlights the need for a more transferable and readily applicable framework for assessing the technological potential of individual scientific papers across fields.Complementary text- and bibliography-based methods may help detect technological relevance before patent citations become observable.

## 2.3. Reflections

The review identifies two challenges: recognizing promising papers within a growing scientific literature and assessing their potential before downstream impact becomes observable. Patent citations are retrospective, gatekeeper approaches capture identifiable actors, and structural or semantic mapping often operates at aggregated levels. These approaches offer limited guidance for early paper-level assessment.

LLMs may help address this gap by detecting semantic signals of technological relevance beyond citations or predefined topic structures. Existing applications focus mainly on academic value; their capacity to assess downstream technological relevance remains underexplored.

Technological impact differs from academic quality: highly cited papers may have limited applicability, while less-cited papers may contain useful methods or materials. We link LLM-based assessments to observed patent citations to examine their potential and limitations as paper-level indicators. This retrospective comparison provides a basis for investigating prospective assessment, rather than establishing real-time forecasting ability.

Finally, there are many generic limitations of patent-based research. In particular, few patents generate commercial value, the extent to which inventions are protected by patents varies between fields, and citations from patents do not necessarily reflect use of the cited research since these citations can be added for many different reasons. Nevertheless, patent citation analysis can help identify some types of technology relevant research.

# 3. Methodology

The research design comprises three stages (Figure 1): collecting and preprocessing paper-level data; defining evaluation dimensions, selecting LLMs, and designing prompts; and comparing LLM scores with observed patent-based impact to assess performance and limitations.

## 3.1. Data source

Assessing technological impact requires linking publications to their later technological uses. Patent citations provide a widely used indicator because they offer observable evidence that scientific knowledge has been incorporated into, or at least is relevant to, technological invention (Hall et al., 2000; Trajtenberg Manuel, 1990). Papers cited more frequently by patents are therefore more likely to have generated technological value (Ke, 2018).

Lens.org was used to construct a large-scale dataset linking scientific papers and patent records. Lens.org integrates bibliographic and patent data, providing publication metadata, together with links between scholarly records and global patents. This integrated structure makes it suitable for tracing science–technology connections at scale and for constructing paper-level measures of real technological impact based on patent citations, which has been widely used in studies of the interaction between science and technology (Jefferson et al., 2018; L. Liu et al., 2023). Although there are alternatives, such as Dimensions.ai and Web of Science with the Derwent Innovation Index, none are clearly superior to Lens.org.

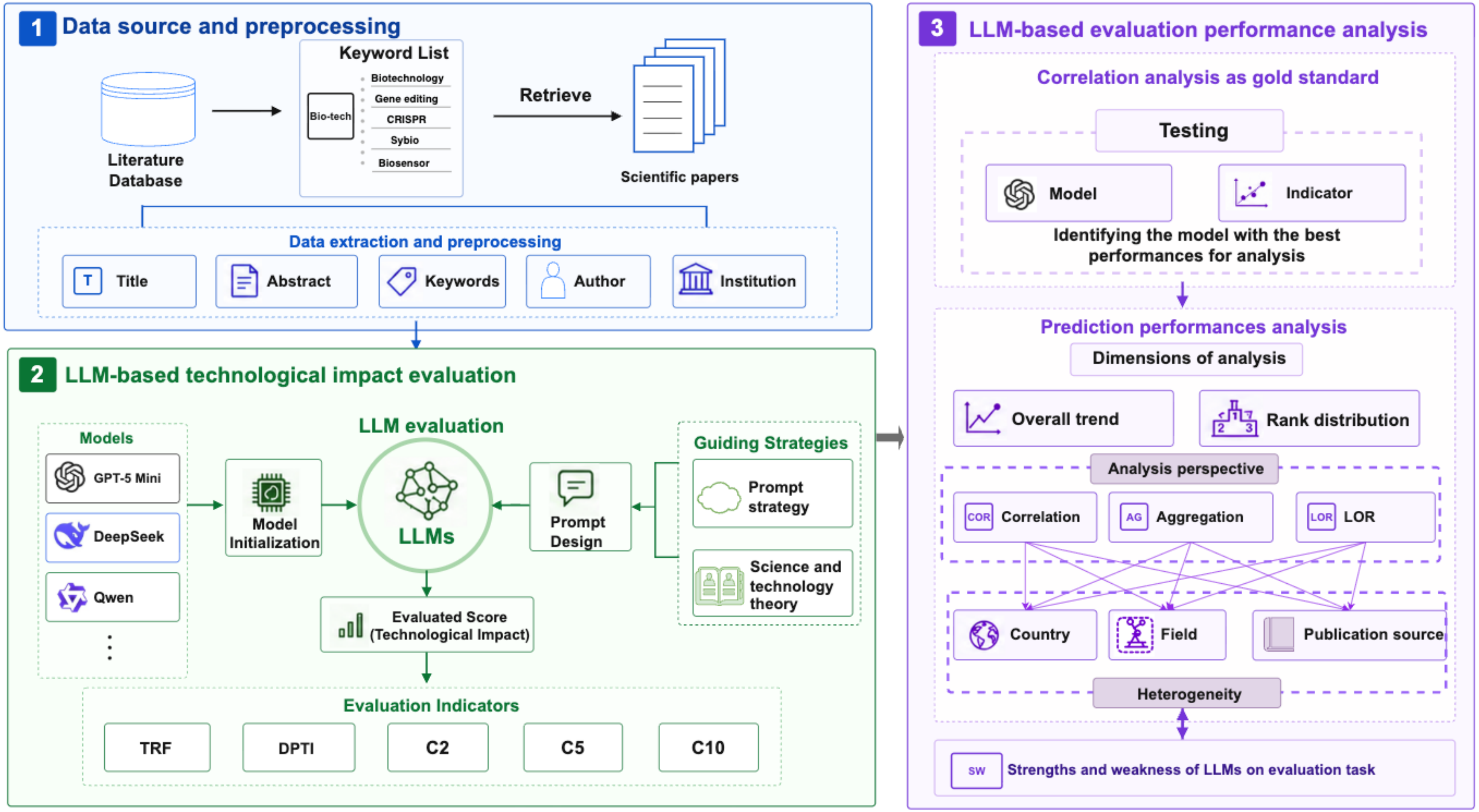

Figure 1. Research design for assessing the ability of LLMs to predict the technological impact of academic papers. Acronyms are explained in the main text.

## 3.2. Data source and preprocessing

Biotechnology was selected as an empirical case for analysis. It provides an appropriate empirical setting for this study because it is widely recognized as a strategic technology area by governments, think tanks, and academic communities (Condoleezza Rice et al., 2025; National Science and Technology Council, 2024). Scientific discoveries in biotechnology are closely connected to downstream technological invention, making this field well suited for examining whether LLMs can evaluate the technological impact of scientific research. Accurately identifying scientific works with substantial technological potential in biotechnology can better support drug R&D by helping firms detect promising knowledge inputs at an early stage, accelerate the translation of basic discoveries into therapeutic applications, and improve the allocation of innovation resources.

We collected biotechnology-related publications from Lens.org on January 24, 2026. To ensure broad coverage, we developed a keyword list for retrieval (see Supplementary Note 1). Because only a small minority of scientific papers receive patent citations, retrieving all biotechnology publications would generate a highly skewed sample dominated by uncited papers. We therefore first collected papers that had received at least one patent citation, yielding 177,125 records.

We restricted the patent-cited sample to 1980–2015, leaving 112,475 records and allowing at least ten years for citations to accrue. Short citation windows can poorly represent long-term impact (J. Wang, 2013), while biotechnology before 1980 differs from its contemporary research structure. To include papers without patent citations, we randomly sampled 20,000 uncited papers published during the same period and merged them with the patent-cited records. Patent-cited papers were fully enumerated within the sampling frame, whereas uncited papers had a lower inclusion probability. Unweighted analysis would therefore underrepresent the latter. We applied inverse probability weighting (IPW) when estimating Pearson and Spearman correlations and assessed sampling uncertainty using bootstrap standard errors and 95% confidence intervals (Supplementary Note 2).

Each publication was assigned a weight equal to the inverse of its inclusion probability. All publications with positive patent citation counts received a weight of 1, whereas the 20,000 publications randomly sampled without replacement from 463,573 uncited publications received a weight of 23.18:

$$w_i = \frac{1}{\pi_i} = \begin{cases} 1 & P_i > 0 \\ \frac{463573}{20000} = 23.18, & P_i = 0, \end{cases} \quad (1)$$

where $\pi_i$ is the inclusion probability and $P_i$ is the patent citation count. These design weights were retained after applying analytical eligibility criteria and excluding records with missing scores, because these exclusions did not alter the original sampling probabilities. The weights address unequal sampling probabilities but do not additionally correct for potential selection associated with missing data.

We next determined what input information to provide to the LLMs. Prior studies suggest that using full texts does not necessarily improve LLM-based evaluation of scientific papers, and may even reduce performance because long and complex texts can introduce noise or distract the model from the core contribution (Thelwall, 2025c). We therefore used titles and abstracts as the main textual inputs; records without abstract information were excluded, leaving 107,615 publications for analysis (see Fig. S2 and Table S1). Prior research on science–technology linkages suggests that patent citations are shaped not only by research content, but also by authors, institutions, funding sources, and publication venues (Bikard, 2018; Ding et al., 2017). We therefore provided the LLMs with author, institutional, funding, and journal information, together with the title and abstract, as the input for technological-impact evaluation.

## 3.3. LLM-based technological impact evaluation

### 3.3.1. Indicators used for technological impact evaluation

We define a set of Technology-Relevance Features (TRF): paper-level characteristics that may indicate whether scientific knowledge can be translated into technological inventions. Conventional criteria used in LLM-based scholarly quality evaluation, such as rigor, originality, and significance, do not directly indicate whether the knowledge produced by the paper can be implemented, reused, embedded in technological systems, or applied to specific industrial problems. The purpose of the TRF is to focus instead on features that are theoretically linked to science-to-technology transformation. The proposed TRF includes five dimensions derived from a literature review of factors that help research to be used in inventions: Technical Implementability (Bryan, 2019), Industry/Application Context Fit (Dorta-González et al., 2024; Tijssen, 2001), Architectural/Interface Position (Ahmadpoor & Jones, 2017; F. Han & Magee, 2018), Modularity and Reusability, and Problem Specificity (Dorta-González et al., 2024) (see Supplementary Note 3). These dimensions assess whether a paper provides an actionable technical method or mechanism, is connected to practical or industrial contexts, occupies a meaningful position within a technological system, offers reusable knowledge components, and addresses a clearly specified technical problem.

To complement the feature-based TRF assessment, we constructed outcome-oriented indicators: Directly Predicted Technological Impact (DPTI), alongside C2, C5, and C10 scores for 2, 5, and 10 years after publication, respectively. This design compares feature identification with direct estimates of downstream patent-based impact.

### 3.3.2. LLM evaluation strategy: models, prompts, and scores

Prior evidence suggests that LLM performance varies across tasks, models, and evaluation settings (Thelwall, 2025f). We therefore compared several cutting-edge models, Qwen-Flash, DeepSeek-V3.2, and GPT-5-mini, with queries submitted on April 13, May 8, and May 21, 2026, respectively, to assess their relative performance in predicting the technological impact of scientific papers.

Prompts followed established evaluation practices and design advice (B. Han et al., 2025; Kojima et al., 2023; Wei et al., 2023) (Supplementary Note 4). Models assigned integer scores of 0–5 to the five TRF dimensions and 0–100 to DPTI, C2, C5, and C10, using titles, publication types and years, authors, fields, keywords, funding, and abstracts.

To reduce scoring variation (Thelwall, 2025a, 2025b), each model scored each indicator three times; analyses used the mean of the three runs.

# 4. Results

## 4.1. LLM predictions of technological innovation or patent citation counts

In the following analysis, we used patent citation counts as the benchmark indicator of technological impact, even though it is imperfect. To reduce the influence of skewness and extreme values in citation distributions, we used the log-transformed measure $ln(patent_citation + 1)$, as the main outcome variable. A stronger association between LLM-based scores and this benchmark indicates better predictive performance.

This section benchmarks the performance of different LLMs and evaluation indicators in patent citation count prediction. We compared three models and three main indicators using Spearman and Pearson correlations. Considering publication years may still affect patent citation counts even after applying a citation-window restriction, we further examined the correlation results using cumulative publication-year cutoffs. Specifically, the correlation reported for a given year is calculated based on all papers published in or before that year (Figure 3).

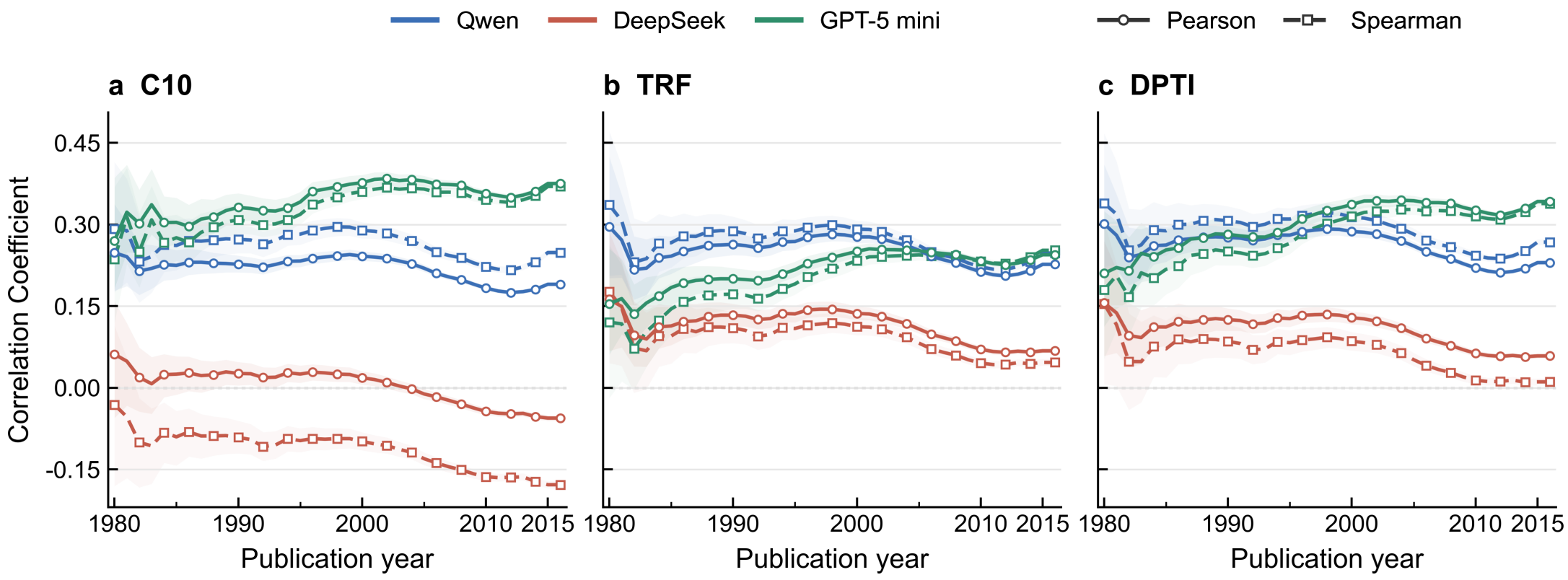

Figure 3. IPW-weighted Pearson and Spearman correlations between three types of LLM-based estimation and logarithm of patent citation between 1980 and 2015 by publication year.

The results show that GPT-5-mini generally has stronger and more stable correlations across publication years although the results produced by Qwen outperform GPT-5-mini to some degree (Figure 2). Among the indicators produced by GPT-5-mini, C10 shows the strongest overall correlations over the time range, followed closely by DPTI, whereas TRF performs relatively weakly.

TRF performs less well than C10 and DPTI, although its results are more stable across models. This may reflect factors beyond the five features, including technological trajectories, industrial demand, patenting practices, absorptive capacity, and researcher or institutional prestige (Sine et al., 2003). The feature-based approach offers interpretability, while direct estimates may capture broader contextual signals. These approaches may therefore serve complementary evaluation needs.

We selected GPT-5-mini and C10 for subsequent analyses because they showed the strongest overall benchmark performance. TRF remains useful for interpreting the content-based sources of technological relevance.

## 4.2. Overall validity of LLM-based indicators

We next examined whether LLM scores capture trends in patent impact (Figure 4) and outperform random selection across impact ranges (Figure 5).

We divided C10, TRF, and DPTI scores into 20 bins (Figure 4) and calculated mean log-transformed patent citations, $ln(patent\ citation + 1)$, within each bin. Higher scores generally correspond to more patent citations. C10 has the strongest overall association; DPTI follows a similar pattern but varies less between intermediate bins. TRF changes little in the middle range and increases toward the upper end. Within-bin Spearman correlations weaken in the highest bin for all three indicators, suggesting limited ability to rank highly rated papers precisely. Cumulative advantage may contribute (Hammarfelt, 2021): visibility to inventors and examiners can generate further citations that textual assessment may not anticipate.

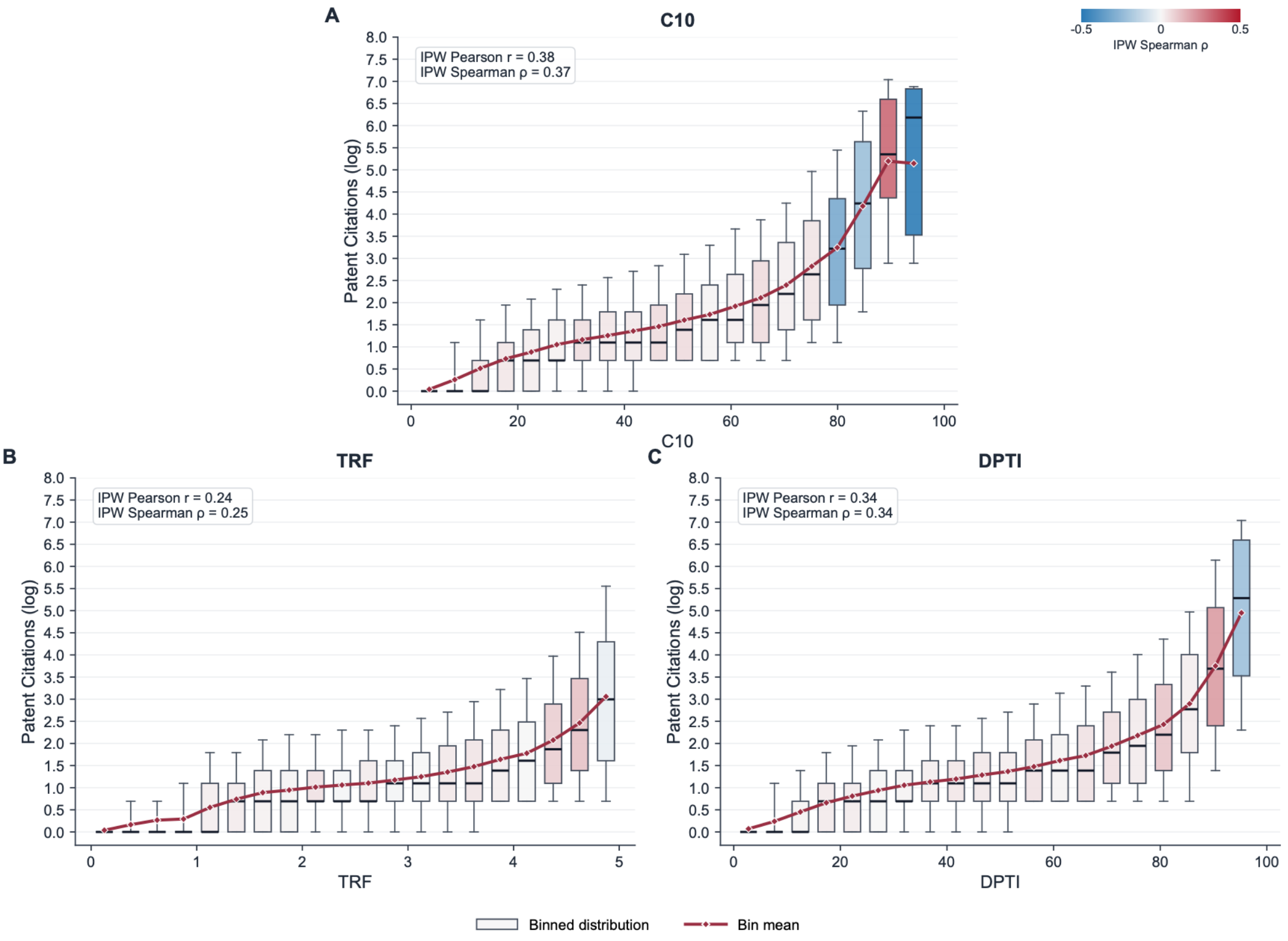

Figure 4. Patent citation impact against three types of binned LLM-evaluated technological impact scores.

We next compared LLM-based screening with random selection (Figure 5). We calculated the distribution of papers across predicted-score third as low, medium and high impact based on estimated and observed technological impact, each divided into three equal-width intervals, and compared it with the random baseline selection. Figure 5 reveals three patterns. First, all three indicators assign relatively high scores to some papers with low observed patent impact where a notable share of papers with low patent impact are assigned to higher predicted-score groups, indicating that LLM-based indicators should not be interpreted as precise paper-level forecasts. Second, the indicators differ in the segments they identify most effectively. TRF and DPTI are stronger at detecting high-impact papers, capturing 89.0% and 72.2% of papers in the top third of the observed-impact distribution, respectively. C10 captures fewer high-impact papers (55.9%) but better differentiates medium-impact papers, correctly classifying 69.7%. Third, recall- and precision-based heatmaps reveal distinct screening profiles. TRF achieves higher recall by assigning high scores more broadly, making it better suited to broad screening. By contrast, C10 is more selective: papers with high C10 scores are substantially more likely to be truly high-impact, reaching 10.6 times the random baseline. This suggests a coverage-precision trade-off: TRF is preferable for broad discovery, whereas C10 is better for identifying a smaller set of candidates for further assessment.

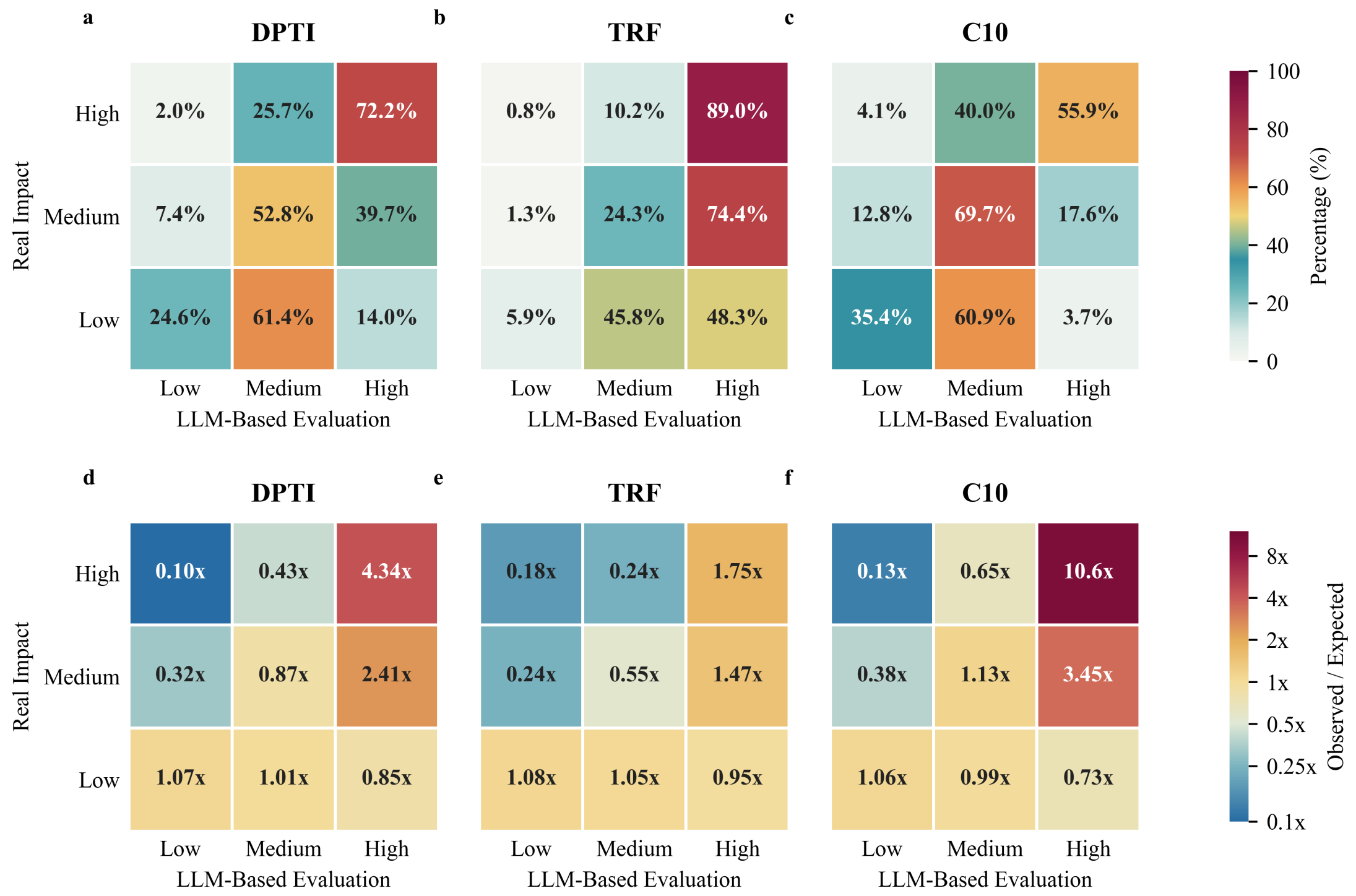

Figure 5. LLM-based identification of high citation impact papers observed hits and lift over random (Note: The upper cells report the observed share of truly identified high citation impact papers. The lower cells report the LOR, indicating how much better the LLM-based evaluation performs than random sampling)

## 4.2.Heterogeneity analysis: research field, publication venue and country

We further examined whether heterogeneity exists in the technological impact evaluation tasks across research fields, publication sources, and publication countries. For each heterogeneity analysis, we focus on two metrics: the Spearman correlation coefficient and Lift

Over Random (LOR). The former captures the overall association between LLM-based scores and real patent citation counts, while the latter evaluates whether LLMs can identify highly impactful papers more effectively than random selection. Together, these two metrics provide more detailed evidence on the strengths, limitations, and boundary conditions of LLM-based technological impact evaluation.

### 4.2.1. Field differences

We matched Lens.org records to OpenAlex using DOI, MAGID, and PMCID to obtain field classifications. We matched 102,559 papers (95.3%); other records lacked identifiers. OpenAlex classifications are assigned using machine-learning models based on titles and abstracts.

**(1)Correlation analysis**

Correlations vary across fields (Figure 5). Higher correlations in Pharmacology (ρ = 0.494) and Physics & Astronomy (ρ = 0.471) may reflect more explicit descriptions of technical problems and applications in titles and abstracts. This explanation remains tentative because semantic visibility was not directly tested.

In contrast, lower correlations are observed in fields where patent citations may reflect contextual or explanatory uses, such as Business & Management (ρ = 0.059), Arts & Humanities (ρ = 0.09), Social Sciences (ρ = 0.113), Economics & Finance (ρ = 0.146) where patent citation counts may be difficult to infer from textual content. For example, the work "The unseen world: reflections on Leeuwenhoek (1677) 'Concerning little animals'" (Arts & Humanities), which has 10 patent citations, is often adopted for contextual or narrative purposes. It carries limited technological implications, making it difficult for LLMs to accurately estimate its true technological impact.

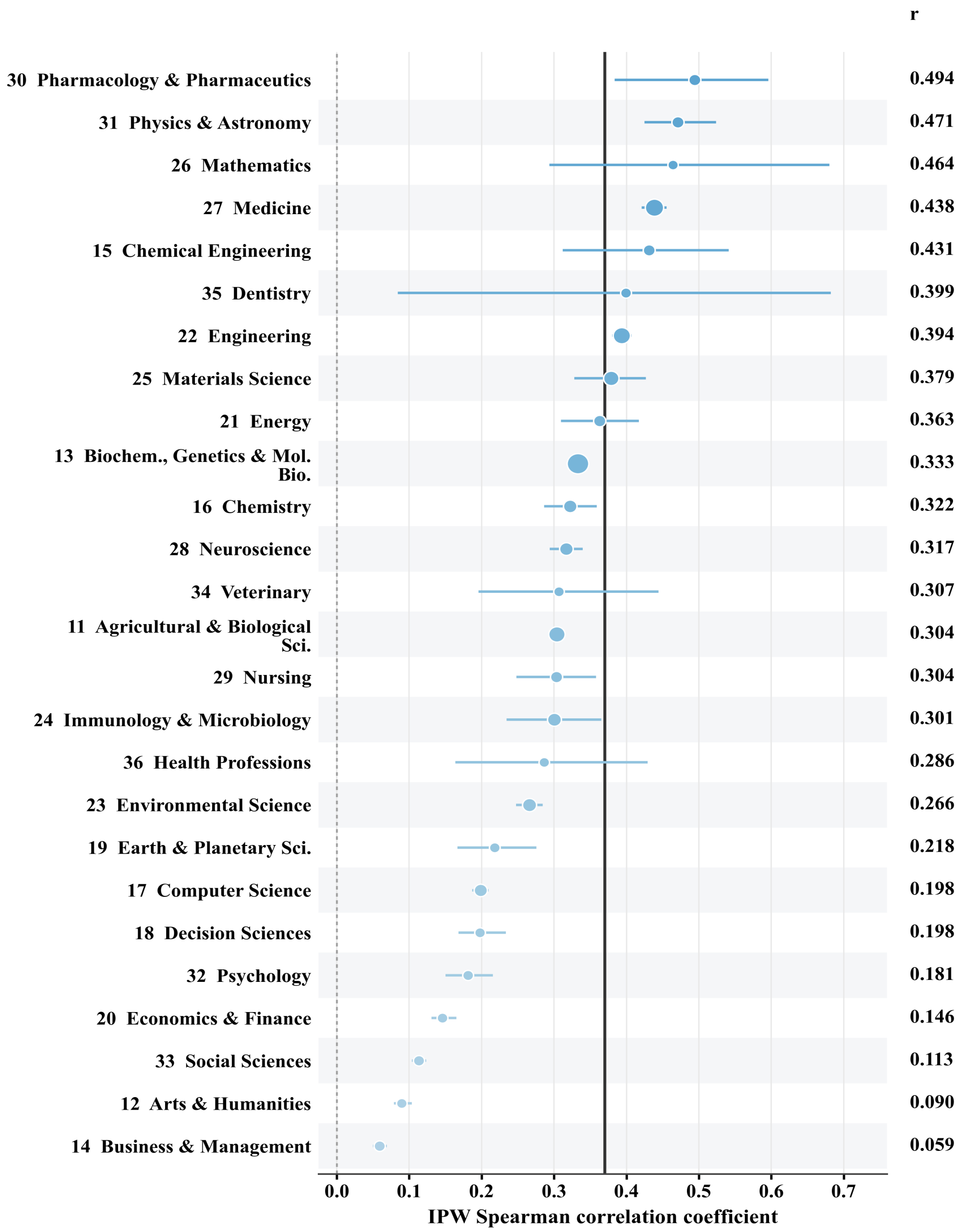

Figure 6. Disciplinary heterogeneity in the Spearman correlations between C10 scores and patent citation counts across fields for the papers in the biotechnology dataset.

### (2) Field-level correlation between citations and C10

We next aggregate paper-level results to the field level to examine whether LLM-based results align with real patent impact. For each field, we calculate the mean C10 score and the mean log-transformed patent citation count. Figure 7 shows a strong positive association between the two measures across fields, with Pearson's $r = 0.808$ ($p < 0.001$) and Spearman's $r = 0.895$ ($p < 0.01$).

These associations indicate aggregate alignment: fields with higher mean C10 scores also have higher mean patent impact. LLM-based indicators may thus help identify areas for further technological monitoring. In this biotechnology sample, materials science, immunology, microbiology, and chemical engineering show strong technological relevance, although this does not establish their importance across all critical technologies.

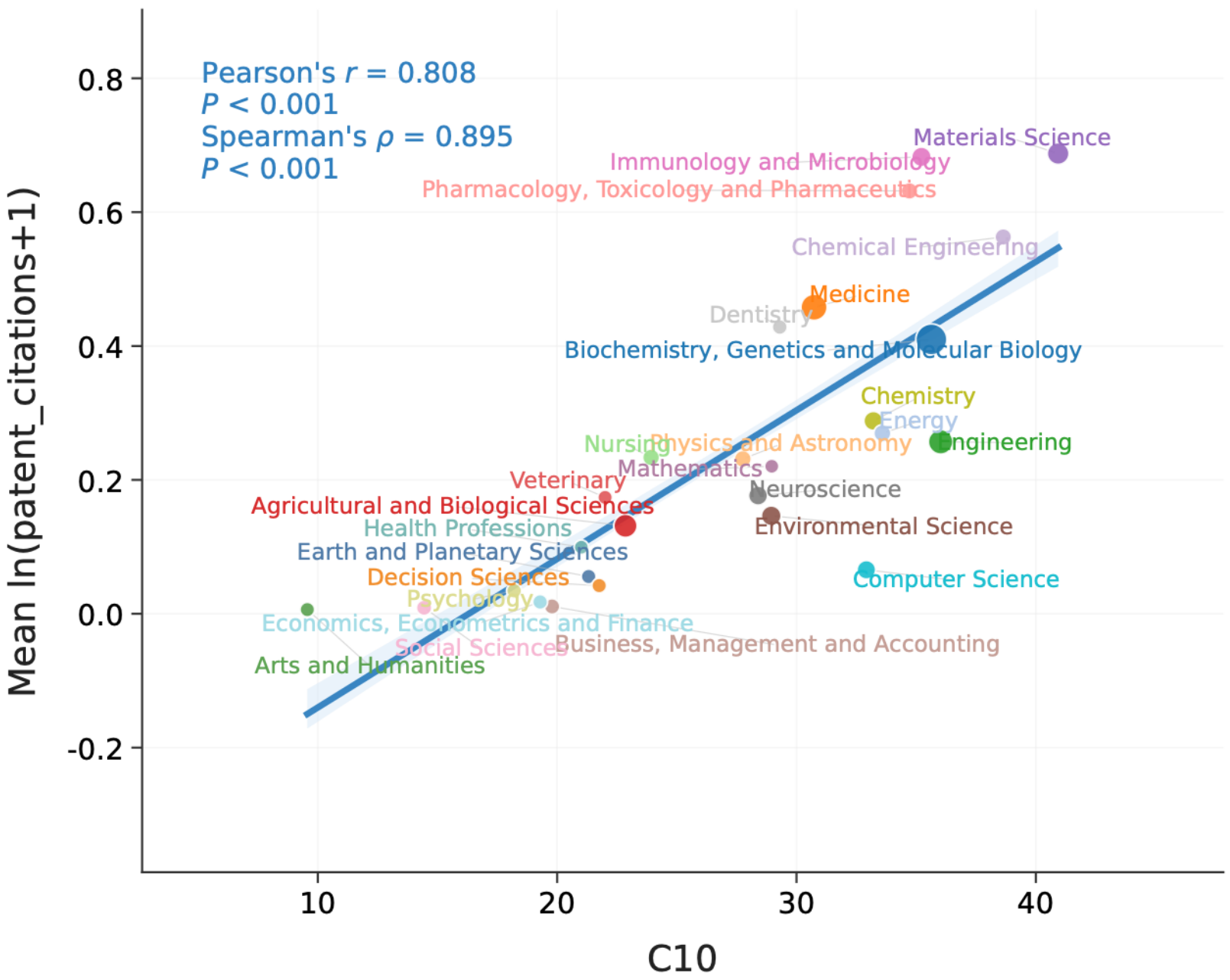

Figure 7. The relationship between mean C10 scores and patent citation counts across fields.

**(3) LOR analysis**

We used LOR to assess whether LLM-based selection identifies highly patent-cited papers more effectively than random selection. Prior estimates suggest that technological value is concentrated in a small share of patents (Feldmeier, 2009; Stryber, 2024), motivating examination of the upper tail of paper-level patent impact.

We define high technological impact papers (HTP) as the top 1% by patent citation count. For each C10 selection threshold, LOR compares the share of HTP among selected papers with the random-selection baseline. Values above one indicate better screening performance than random selection.

All fields exceed the random baseline (Figure 8), but lift varies across fields. It is highest when selecting the top 1–40% of C10-ranked papers and declines as selection expands. This supports selective screening for highly patent-relevant papers, with more limited discrimination across the full impact distribution.

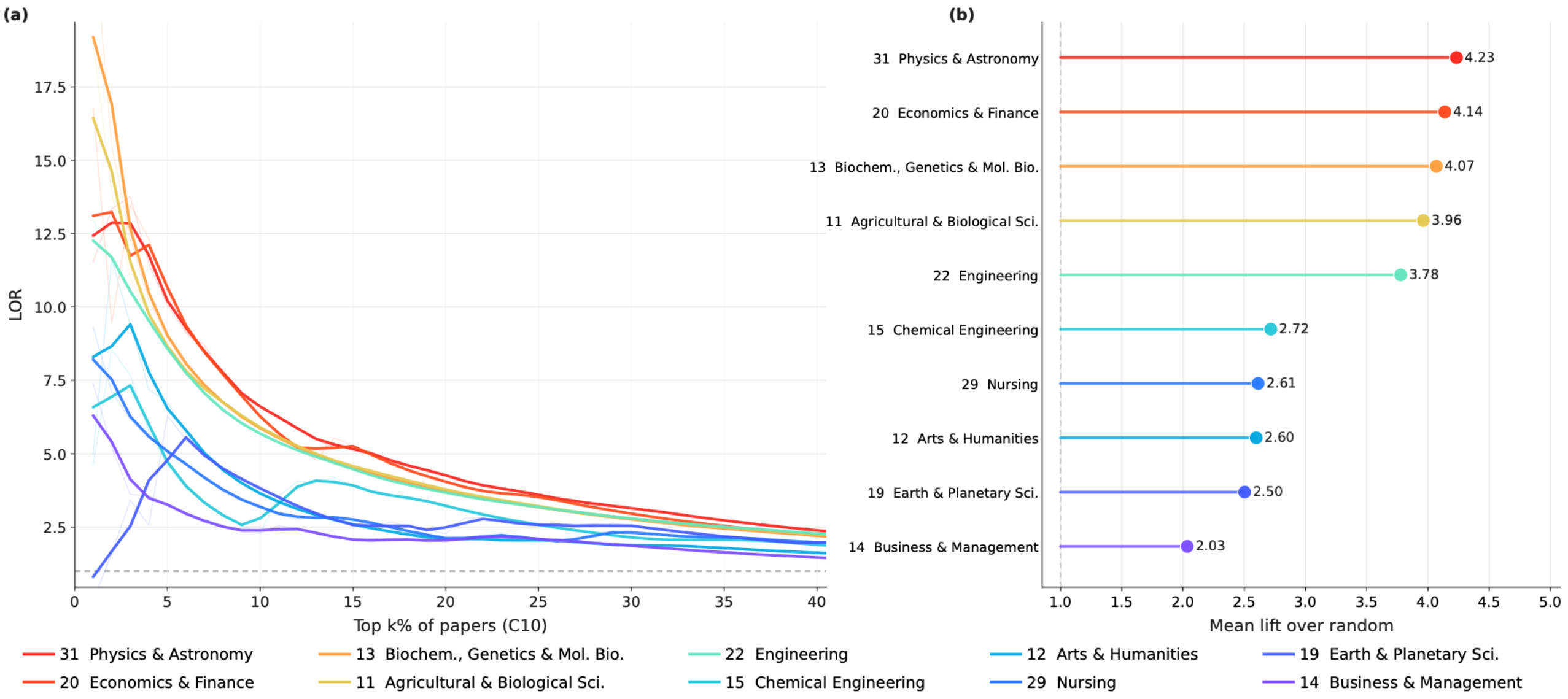

Figure 8. Field heterogeneity in LLM-based identification of patent-relevant biomedical papers compared to the random selection.

### 4.2.2. Publication venues

#### (1) Correlation analysis

We further examined heterogeneity across publication venues. Because the dataset includes many sources, we report the ten sources with the highest and lowest Pearson correlations between C10 and patent citation counts.

As shown in Figure 9, the highest correlation is observed in Industrial & Engineering Chemistry Research (r = 0.897, n = 33), Cell Stem Cell (r=0.884, n = 48), and Science Translational Medicine (r =0.861, n = 61), whereas the lowest appears in Endocrinology (r = −0.537, n = 66), RSC Adv (r=-0.52, n = 40), Organic & Biomolecular Chemistry (r=-0.436, n = 88). These results identify publication venue as an important boundary condition for LLM-based technological-impact evaluation, suggesting that LLM-based scores should not be treated as equally reliable across all sources. Instead, their validity depends on whether the journal context makes technological relevance visible in the paper-level information available to the model.

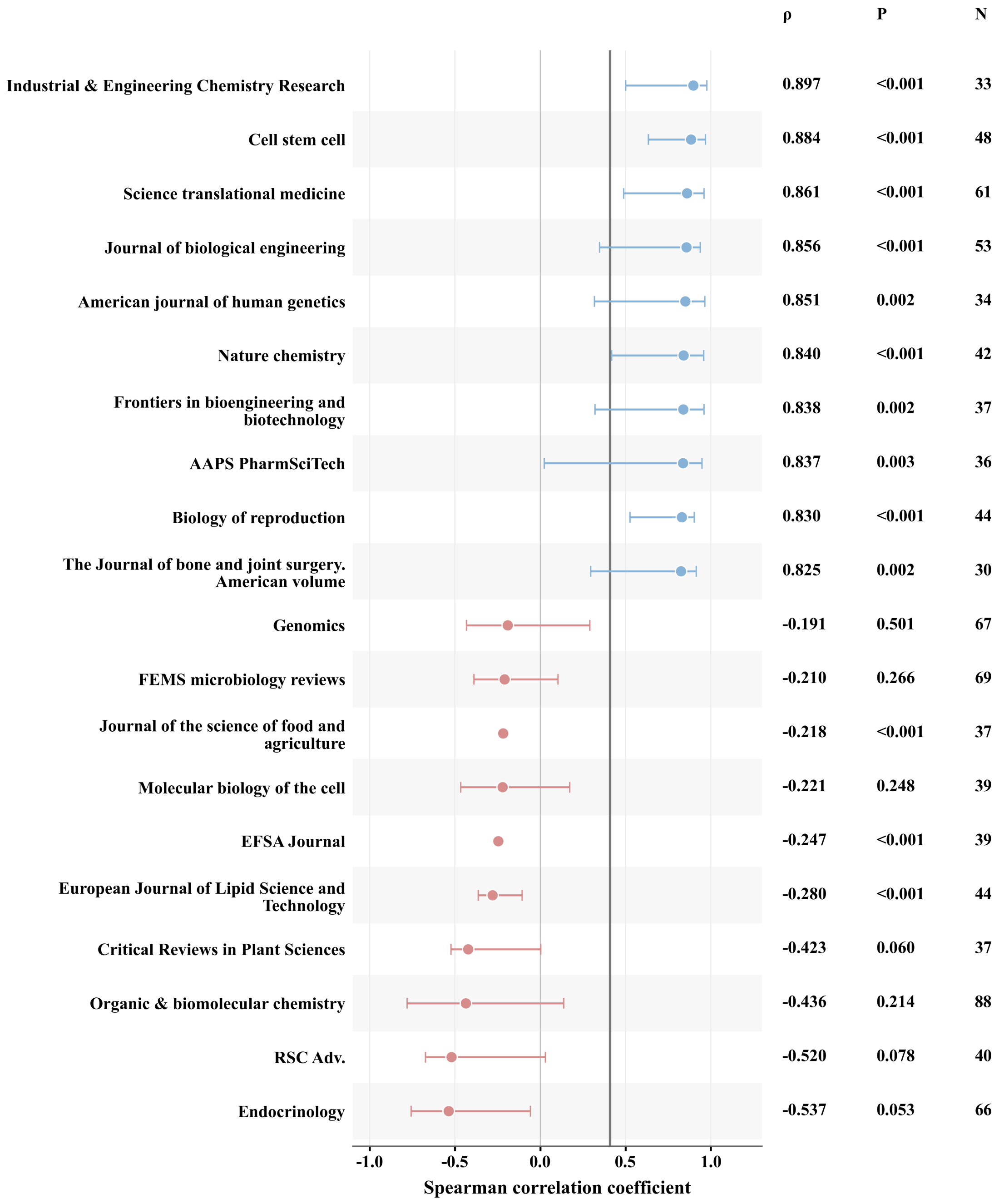

Figure 9. Publication sources heterogeneity in the association between GPT-5-mini C10 scores and patent citation counts.

### (2) Aggregation analysis

We next aggregate the results to the publication-source level and examine whether source-level mean C10 scores align with real patent impact. For each publication source, we calculate the mean C10 score and the mean log-transformed patent citation count.

The association remains positive and statistically significant at the publication-source level, with Pearson's $r = 0.602$ ($p < 0.001$) and Spearman's $\rho = 0.572$ (Figure 10). Venues (mainly journals) with higher average LLM-evaluated C10 also tend to receive more patent citations on average. However, the publication-source-level association is weaker and more dispersed than the fields aggregation. This indicates that journal-level technological impact is shaped not only by the scientific content of papers, but may also depend on source-specific factors such as journal scope, field composition, and connections to applied research communities.

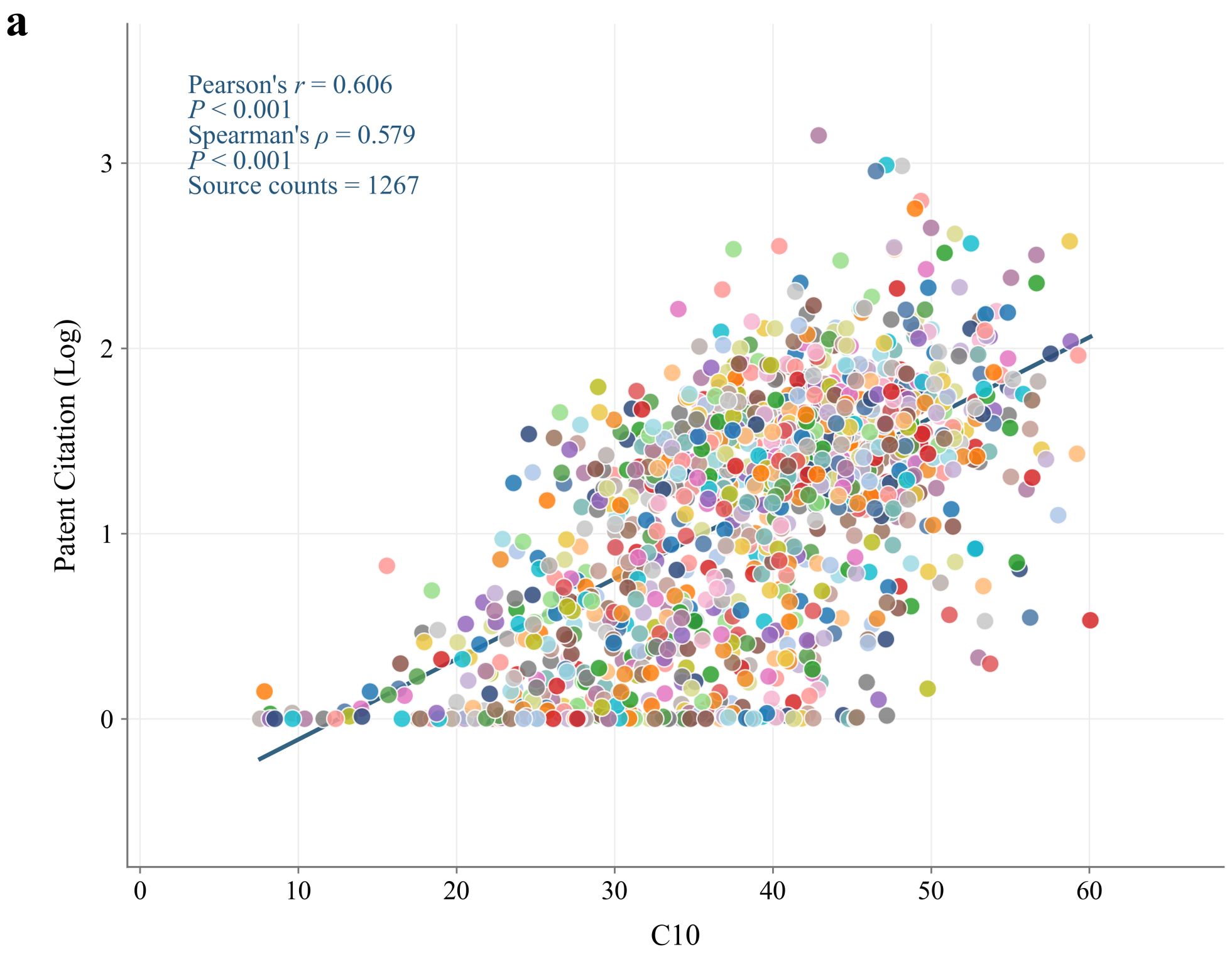

Figure 10. Logged patent citation counts against LLM-based technological impact estimates aggregated by source titles (mainly journals).

The publication-source analysis also reveals substantial discrepancies between LLM-based scores and observed patent impact. Several sources, including Lawrence Berkeley National Laboratory (n = 20), Chinese Bulletin of Life Sciences (n = 21), Progress in Biochemistry and Biophysics (n = 13), have almost no patent citations but receive relatively high C10 scores (>40). This pattern suggests that the LLMs may be sensitive to technical vocabulary and application-oriented rhetoric, treating them as signs of technological potential. It may also reflect international differences in the use of patents. However, empirical evidence shows that patent citations capture a narrower institutionalized form of science–technology linkage: knowledge must be visible, citable, codified, and useful within invention and patenting processes (Narin et al., 1997). Moreover, patent citations are an imperfect proxy for knowledge flows, shaped by examiner practices, source visibility, and citation conventions (Roach & Cohen, 2013). In contrast, the three journals with the highest observed patent-based impact, including Bio/Technology (Nature Biotechnology; n = 51), Medical Engineering & Physics (n = 11), and Annual Review of Biochemistry (n = 56), showed relatively low C10 scores despite high patent citation counts. This discrepancy may partly reflect citation advantages associated with journal visibility and prestige, alongside the knowledge-synthesis function of review articles (Yamashita, 2018). These factors may influence patent citation counts in ways that are not fully captured by C10 scores. For Medical Engineering & Physics, the small sample size may also make the estimated journal-level impact sensitive to a few highly cited papers.

These outliers suggest that perceived applicability and observed patent impact can diverge. Unobserved uptake mechanisms and source visibility may contribute, but the present analysis cannot determine whether discrepancies reflect model error or technological potential missed by patent citations.

### (3) LOR analysis

Figure 11 shows substantial venue-level variation. The strongest lift appears in *Cell* (n = 272), where LLM-based selection performs 9.14 times better than random selection, followed by *Nucleic Acids Research* (n=1,147), *Nature Biotechnology* (n=509), *Science* (n = 674) and Nature (n = 871). In contrast, *The FASEB Journal* (n = 113) shows the lowest LOR, with performance below the random baseline.

In sources such as *Nature Biotechnology* (n=509) and *Science* (n=674), C10 is especially effective at identifying HTP papers among the most highly ranked outputs. For example, in *Nature Biotechnology*, the LOR exceeds 40 within the top 1% of predicted papers, indicating strong enrichment relative to random selection. By contrast, in sources such as *The FASEB Journal* (n = 113) and *SPIE Proceedings* (n = 67), the lift remains close to, or below, the random baseline. This suggests that LLM-based indicators should be used with venue-specific calibration rather than applied as a universal screening rule across all publication sources.

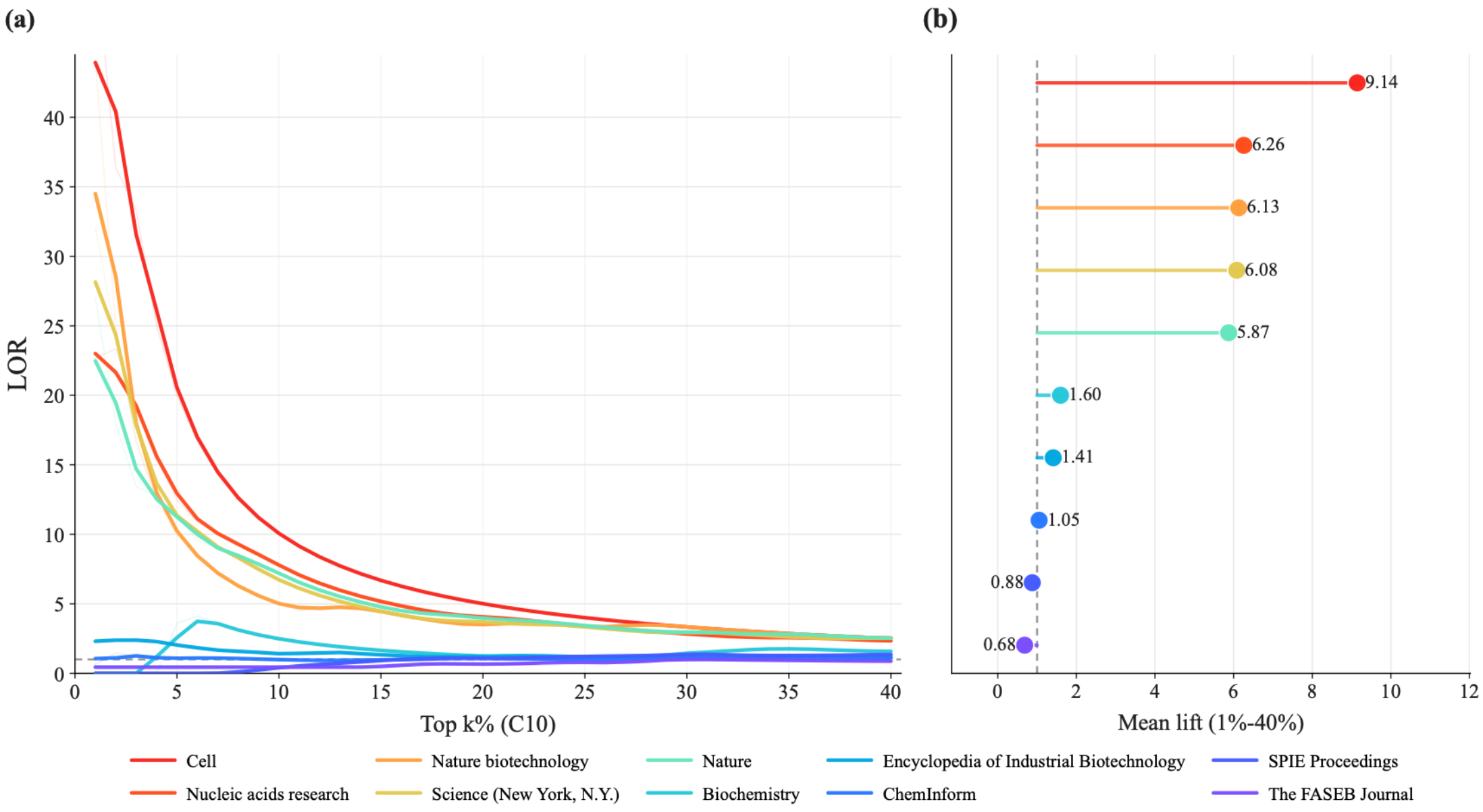

Figure 11. LOR Performance across publication sources for scientific paper technological impact evaluation

### 4.2.3. Variation across Countries

Finally, we examined the heterogeneity across national contexts. We identified the source country of each paper using the metadata provided by Lens.org and assessed country-level heterogeneity from three perspectives: correlation, aggregation, and LOR.

### (1) Correlation analysis

To reduce instability caused by small samples, we retained only countries with more than 20 publications. As shown in Figure 12, the correlation between C10 and patent citation counts varies substantially across countries. Higher correlations are observed in Denmark (r = 0.818, n = 28), United Kingdom (r = 0.557, n = 23,840), New Zealand (r = 0.517, n = 168), whereas lower correlations appear in Ukraine (r = 0.033, n = 37), Romania (r = 0.04, n = 24), and Belgium (r= 0.053, n = 55).

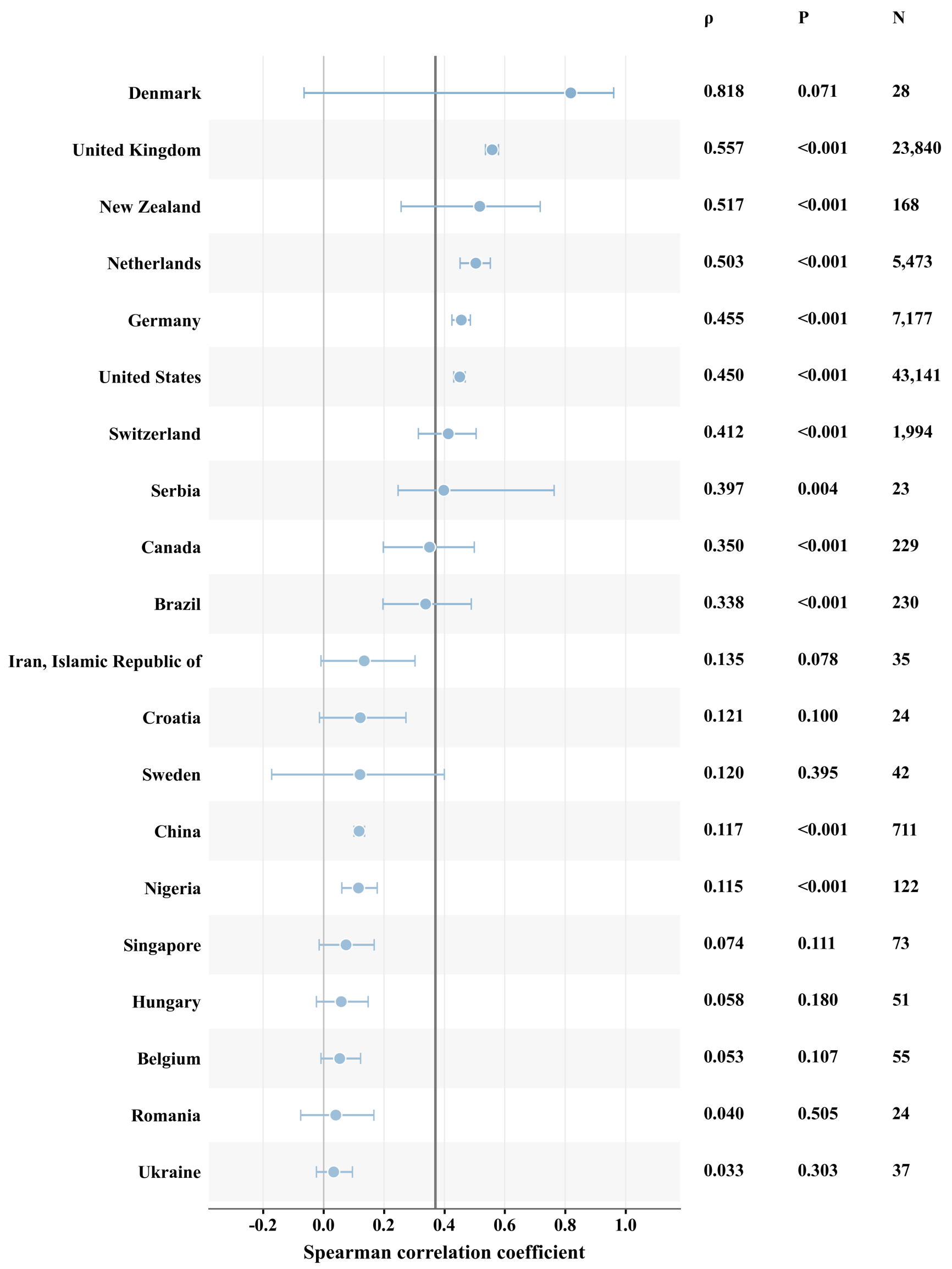

Figure 12. Country-level heterogeneity in the relationship between LLM scores and patent citation counts (10 countries with highest and lowest correlation coefficient)

## (2) Aggregation analysis

Across 49 countries, mean C10 and mean log-transformed patent citations are moderately associated (Pearson's r = 0.576, p < 0.001; Spearman's ρ = 0.566, p < 0.001; Figure 13). Malaysia (n = 10), China (n = 711), and Singapore (n = 73) have comparatively high assessed potential relative to observed impact. Possible explanations include industrial-capability constraints in Malaysia (Chandran et al., 2014), university–industry linkages in China (Tang & Du, 2016), and limited absorptive capacity among some Singaporean SMEs (Ho et al., 2016). These mechanisms were not tested here.

Denmark (n = 28) and the United Arab Emirates (n = 937) show higher patent-based impact than their C10 scores suggest. In Denmark, pharmaceutical capabilities and university connections (Laursen, 1996), together with scientific and experience-based learning (Jensen et al., 2007), may support uptake that paper-level assessments miss.

In the UAE, international collaboration accounted for approximately 77% of medical publications in 2007–2016 (El Rassi et al., 2018), alongside substantial researcher inflows (El-Ouahi et al., 2021). Arts et al. (2026) show that interpersonal connections can facilitate corporate uptake when recipient inventors have relevant expertise. Such networks may extend research visibility to industrial users abroad, although their role in the present results remains untested.

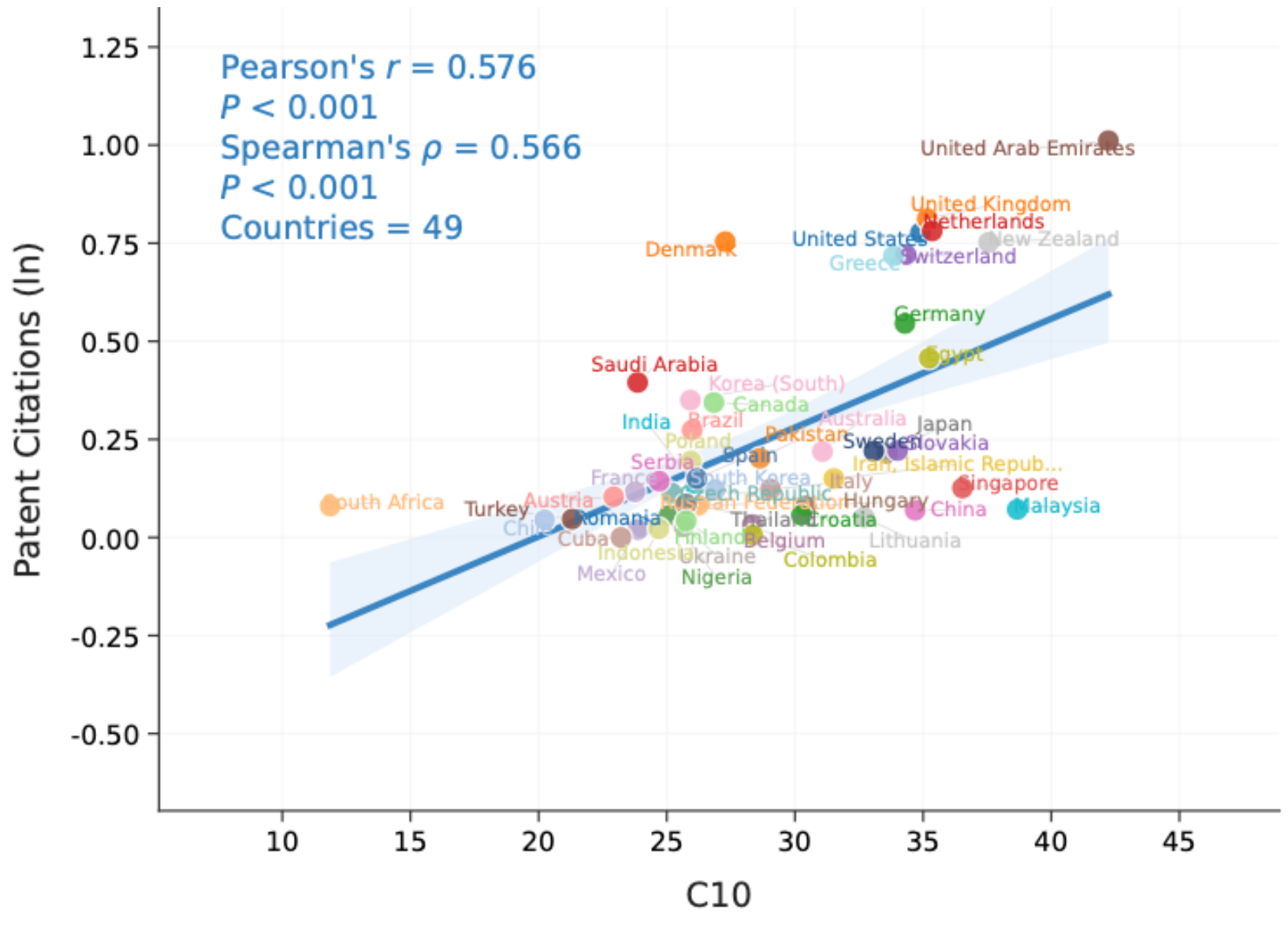

Figure 13. Association between GPT-5-mini C10 scores and logged patent citation counts across countries (n ≥ 10)

These differences suggest that assessed potential and observed patent impact depend on national context. Papers may inform technology without receiving patent citations, while English proficiency may also be associated with evaluation performance (Figure S2). Country-level associations do not establish that language proficiency or innovation-system conditions cause the observed differences.

## (3) LOR analysis

We restricted country-level LOR analysis to countries with more than 200 publications to reduce small-sample instability.

Figure 14b shows that C10 has positive screening power across countries, but the magnitude of this advantage varies substantially. All selected countries exceed the random baseline when identifying HTP papers. The strongest lift appears in the United Kingdom (n = 23,782), India (332), the United States (43,141), France (210), and Switzerland (1,994). In contrast, China (711), South Korea (334), Japan (764), United Arab Emirates (937) and Canada (229) have lower LOR values, although their LORs remain above one.

Lift generally peaks under selective screening and declines toward the random baseline as thresholds expand (Figure 14a). At the top 1% threshold, most countries outperform random selection, with India showing LOR above 10. C10 therefore appears more useful for selecting highly patent-relevant candidates than for ranking full national publication sets.

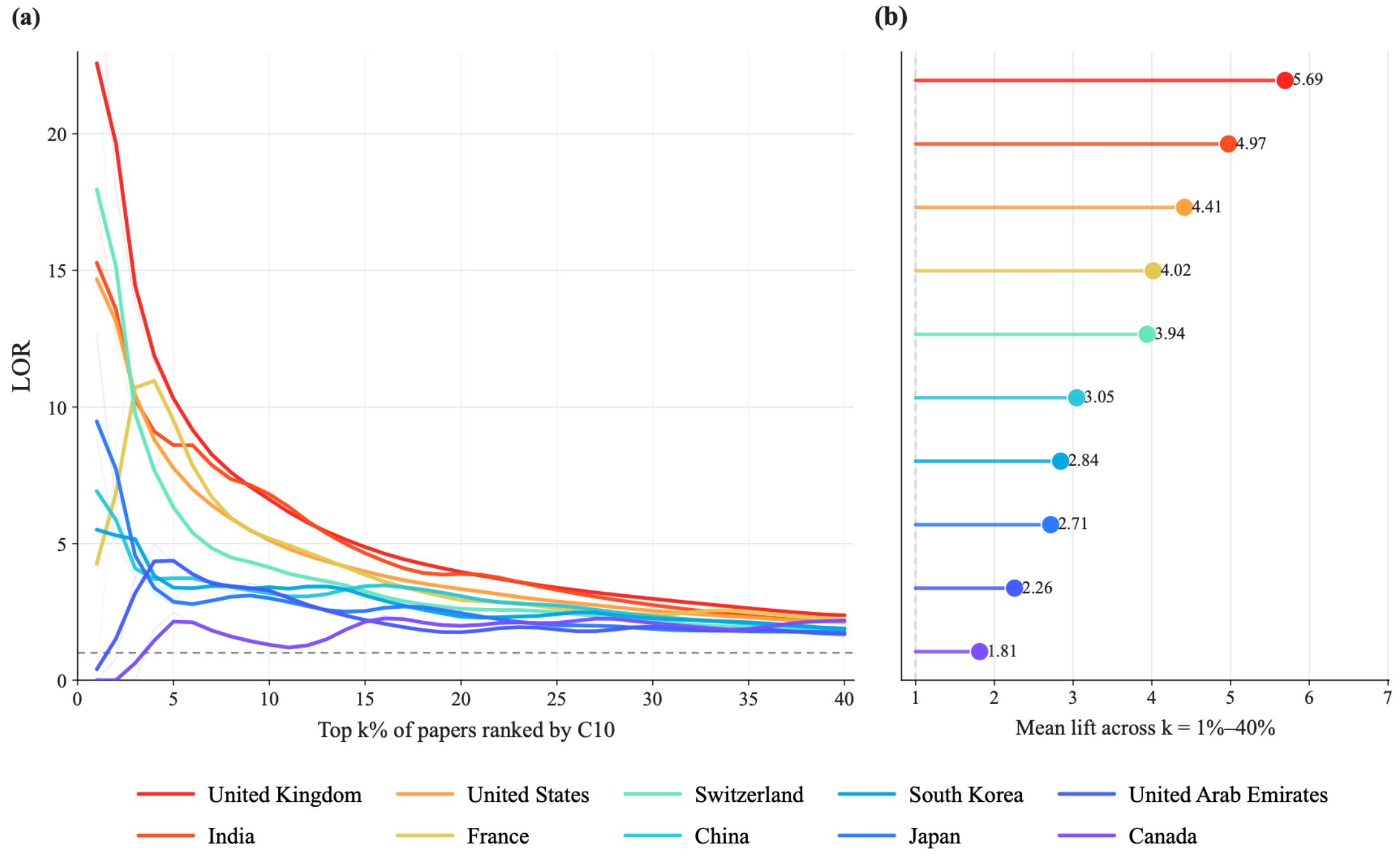

Figure 14. LOR Performance across countries for scientific paper technological impact Evaluation

## 5. Discussion

### 5.1.Summary

This study systematically assesses whether LLMs can evaluate the technological impact of scientific research, using patent citation counts as a proxy for actual technological impact. We constructed a large-scale paper-patent dataset from Lens.org and compared three LLMs: GPT-5-mini, DeepSeek-V3.2, and Qwen-Flash. We used a literature-informed TRF indicator and four outcome-oriented indicators: DPTI, C2, C5, and C10. In the retrospective benchmark analysis, GPT-5-mini has the strongest and most stable associations with observed patent impact, and C10 is the most effective indicator used in the subsequent analyses.

The analysis yields three main findings. First, LLMs have meaningful but limited criterion validity. Their scores are positively associated with patent citation counts and broadly follow the distribution of patent-citation outcomes. LLMs can therefore capture signals of technological relevance in scientific text and bibliographic information. However, they are better suited to identifying broad patterns and supporting field-level monitoring than to estimating the exact technological impact (at least as reflected by patent citation counts) of individual papers.

Second, LLMs are most powerful at identifying papers associated with the highest patent citation counts. Relative to random selection, LLM-based indicators identify more papers in the upper tail of the patent-impact distribution, especially under narrow selection thresholds such as the top 1%. They are less effective at distinguishing low- and medium-cited papers from others. Their main value may therefore lie in helping researchers, firms, and policymakers narrow a large publication set to a smaller candidate pool, rather than in producing a comprehensive ranking of all papers.

Third, LLM-based evaluation is highly context dependent. Criterion validity varies across fields, publication sources, and national contexts. These differences indicate that technological relevance, at least as reflected by patent citation counts, is not equally visible in all textual and institutional settings.

## 5.2.Implications and limitations

This study extends the emerging literature on LLM-based research evaluation in two respects. Prior work has primarily asked whether LLMs can reproduce judgments of academic quality or assist peer review (Liang et al., 2024; Thelwall, 2025e; Thelwall & Yang, 2025; Zhang, Zhang, et al., 2025). Our findings show that they can also recover signals associated with a different evaluative object: the downstream technological relevance of scientific papers, as reflected by patent citations. The results, nevertheless, qualify any claim of general evaluative competence. Positive paper-level associations co-exist with much stronger alignment after aggregation, including a field-level correlation of 0.895, and with the clearest screening gains among the top 1% of papers. LLMs, therefore, appear better at detecting broad and upper-tail patterns than at estimating the exact technological impact of an individual paper. This finding reinforces prior evidence that LLM performance varies across tasks and objects (Thelwall & Kurt, 2025), while suggesting a theoretical explanation for that variation. Evaluation is more reliable when the target construct leaves visible semantic or bibliographic traces in the information supplied to the model. It becomes less reliable when the construct depends on processes that remain latent at the paper level. LLM-based evaluation should thus be understood as an interaction among the evaluation target, its textual representation, and the context in which impact is produced, rather than as a stable capability of the model alone.

The contrast between the theory-informed TRF indicator and the more general C10 indicator further clarifies the science–technology relationship reviewed above. Linear accounts position science as an upstream knowledge base for later invention (Vannevar, 1945), whereas Pasteur's Quadrant, interactive models, and innovation-system perspectives emphasize that translation depends on feedback, use contexts, organizational capabilities, and institutional arrangements (Stokes, Donald E, 2011). TRF operationalizes the content side of this process by assessing implementability, application fit, system position, reusability, and problem specificity. Its positive but weaker association with patent citations indicates that these attributes describe technological affordance: they help explain how a paper could become useful to technology. The pronounced variation across fields, journals, and countries shows, however, that affordance does not guarantee realization. Industrial demand, absorptive capacity, visibility, and patenting practices can influence whether a scientifically promising idea enters an invention process and becomes observable in patent citations. C10 might perform better partly because it can draw on broader contextual regularities, but this advantage comes with lower interpretability and greater exposure to learned salience, language, and prestige biases. In addition, C10 might also exploit funding information, with industry funding encouraging it to give higher C10 estimates. TRF and C10 should, therefore, be treated as complementary measures: the former estimates the content-based sources of technological potential, whereas the latter offers a more effective, though less transparent, screening signal.

These findings also contribute to the measurement literature on science-technology linkages. Patent citations provide direct but retrospective evidence of realized potential knowledge flows (Roach & Cohen, 2013). Gatekeeper measures identify human and organizational bridges but cover only a limited set of actors (Li et al., 2020), while topic or semantic approaches capture broader connections but often operate above the level of individual papers (Xu et al., 2020). LLM-based scores address part of this blind spot by constructing a paper-level, content-based indicator that does not require a citation to have already been observed. They should not, however, be interpreted as a replacement for patent-based measures. The two approaches capture different stages of the translation process: an LLM score reflects the textual plausibility of technological use, whereas a patent citation records an institutionalized connection to invention. Disagreement between them is consequently informative. For example, a high LLM score paired with few patent citations may represent model overestimation, but it may also indicate unrealized or poorly visible technological potential. The reverse pattern may reveal impact driven by contextual factors that are not evident in the paper. Triangulating LLM scores with patent, network, and institutional indicators would therefore provide a more complete account than selecting any single measure. Moreover, because the present analysis validates scores retrospectively, prospective and temporally isolated tests are needed before the method can be claimed to forecast technological impact in real time.

The practical implication is a staged and context-calibrated form of decision support. Research funders, universities, firms, and technology-transfer offices could use C10 to reduce a large publication set to a small upper-tail candidate pool, use TRF to explain why particular papers appear technologically promising, and then rely on domain experts to assess feasibility, demand, intellectual-property conditions, and organizational fit. A universal threshold would be inappropriate: the large differences across fields and publication venues, together with the positive association between English proficiency and country-level correlation performance ($r = 0.355$, $p = 0.029$), call for field-, venue-, country-, and language-specific validation.

Finally, this study has several limitations. First, although patent citations are used as the benchmark for observed technological impact, they do not fully capture the broader technological value of scientific research: Papers can have technological impact in ways that do not lead to patenting. Patent-citation outcomes are shaped not only by the intrinsic technological relevance of papers, but also by patenting strategies, examiner practices, and institutional citation norms. Future studies could incorporate expert-based assessments as complementary validation metrics. Second, this study focuses only on biotechnology, a field with relatively strong science–technology linkages. Generalization to other disciplines requires further empirical validation. Third, the LLM-based evaluation design remains limited. This study compares only three models, and future work could further examine the effects of model selection, prompt design, input content, and scoring strategies to improve the robustness of LLM-based technological-impact assessment. Finally, LLMs may draw on prior knowledge of the papers' technological impact, particularly for prominent publications. Such information leakage could inflate retrospective performance and limit applicability to newly published papers.

## Generative AI usage statement

GPT-5 was used to assist with the writing of selected code and the polishing of portions of the manuscript language. We take full responsibility for the content refined with AI assistance and have conducted a thorough verification of its accuracy.

## Code and data availability

Data and code will be made available on request.

## Acknowledgments

The authors are supported by the Major Special Project for Philosophy and Social Sciences Research of the Ministry of Education (#2025JZDZ097).